\documentstyle[12pt]{article}
\begin{document}
\begin{titlepage}
\mbox{}
\begin{flushright}
UM-P-95/39\\
RCHEP-95/12 \\
April, 1995
\end{flushright}

\vspace{1in}

\begin{center} {\Large\bf Quantisation without Witten Anomalies}\\

\vspace{0.3in}

T D Kieu,\\School of Physics,\\Research Centre for High Energy Physics,\\
University of Melbourne,\\Parkville Vic 3052,\\Australia

\vspace{0.3in}

{\bf Abstract}
\end{center}
\begin{quotation}
\noindent It is argued that the
gauge anomalies are only the artefacts of quantum field theory when certain
subtleties are not taken into account. With the Berry's phase needed to
satisfy certain boundary conditions of the generating path integral, the
gauge anomalies associated with homotopically nontrivial gauge
transformations are shown explicitly to be eliminated, without any extra
quantum fields introduced.  This is in contra-distinction to other
quantisations of `anomalous' gauge theory where extra, new fields are
introduced to explicitly cancel the anomalies.
\end{quotation}
\end{titlepage}
\noindent In a seminal paper~\cite{witten}, Witten has
demonstrated a kind of anomalies of the homotopically non-trivial (large)
transformations of the chiral gauge groups $SU(2)$ and $Sp(N)$, whose fourth
homotopy groups are non-trivial.  The arguments are based on the observation
that there is an odd number of pairs of oppositely signed eigenvalues of the
four-dimensional euclidean Dirac operator flipping their signs as the gauge
fields are moved to non-trivially transformed configurations.  While this has
no observable effect on the generating functional for the Dirac fermions, it
renders the generating functional for an odd number of chiral Weyl fermions
inconsistent.  This kind of anomalies is called Witten anomalies as distinct
from the Adler-Bell-Jackiw (ABJ) anomalies associated with small
(homotopically trivial) gauge transformations~\cite{jackiw1}.

The Witten anomalies can also be demonstrated by embedding the $SU(2)$ into a
bigger group, like $SU(3)$, having trivial homotopy group so that a
connection to the ABJ anomalies can be made and their mechanism can be
employed to demonstrate the Witten anomalies restricted to the
subgroup~\cite{elitzur}. Alternatively, they have also been reproduced in the
hamiltonian approach~\cite{sonoda}.

Notwithstanding all the above, we will argue in this letter that with a
quantisation of chiral fermions~\cite{tdk-berry,tdk-ordered,danos} there is
no gauge anomalies associated with large gauge transformations. The absence
of ABJ gauge anomalies in this quantisation has been
demonstrated~\cite{tdk-ordered} and will be substantiated
elsewhere~\cite{tdk-inv}.

We will not question the derivation in~\cite{witten,elitzur,sonoda} of the
anomalies but will start from where the ref~\cite{tdk-berry} left off, that
is, at the level of the path-integral representation of the generating
functional. We have shown in~\cite{tdk-berry} that work that for fermionic
fields in the presence of external gauge fields the action in the path
integral is the sum of the usual classical action {\it and} an extra
contribution of order $O(\hbar)$.  In other words, the generating functional
is the product of the fermion determinant {\it and} an extra factor.  The
extra contribution comes from the Berry's phase~\cite{berry} of the
one-particle equation of motion.  It is shown in this work that the extra
factor reverses the sign change of the fermion determinant, keeping the
generating functional invariant under large gauge transformations.  The proof
realises the anticipation stated in~\cite{tdk-berry}.

We first consider the four-dimensional Dirac operator in Minkowski metric and
in the temporal gauge, $A_0=0$,
\begin{eqnarray} {\cal D} &=& i\partial_t +
H(t),\label{3.0}
\end{eqnarray}
and
\begin{eqnarray} H(t) &=& -\sum_i
\gamma_0\gamma_i[i\partial_i + A_i({\bf x},t)],
\end{eqnarray} for a finite
time interval $t\in [-T,T]$, where $H(-T) = H(T)$.  The limit $T\to\infty$
will be taken later on.

We denote $E_r(t)$ and $\langle {\bf x}|r;t\rangle$ the eigen-energy and the
{\it single-valued} eigenstate, respectively, of the Dirac fermion
hamiltonian, $H(t)$, which is time-dependent through its dependence on the
time-dependent background fields,
\begin{eqnarray}
H(t) \langle{\bf
x}|r;t\rangle &=& E_r(t) \langle{\bf x}|r;t\rangle .
\label{3.4}
\end{eqnarray}
We assume that for all $t$ the eigenstates are non-degenerate
and the eigen-energies do not cross zero. In anti-commuting $\gamma_5$ with
the operator $\gamma_0 H(t)$, which has the same spectrum as $H(t)$, it is
seen that the eigen-energy spectrum contains oppositely signed pairs,
$\{E_r(t),-E_r(t)\}$.

The zero modes of the Dirac operator, $f_r({\bf x},t)$, with
\begin{eqnarray}
f_r({\bf x},T) &=& \exp\left\{-i\alpha_r\right\} f_r({\bf x},-T)
\label{3.4a}
\end{eqnarray}
is characterised by the Floquet indices
$\alpha_r$~\cite{floquet,niemi}, and is related to the above eigenstate in
the adiabatic approximation as
\begin{eqnarray} f_r({\bf x},t) &=&
\exp\left\{i\int_{-T}^t d\tau E_r(\tau) + i\gamma_r(-T\to
t)\right\}\langle{\bf x}|r;t\rangle. \label{3.5} \end{eqnarray} $\gamma_r(t)$
is the Berry's phase~\cite{berry},
\begin{eqnarray} \gamma_r(-T\to t) &=&
i\int_{-T}^t d\tau \langle r;\tau|\frac{d}{d\tau}|r;\tau\rangle. \label{3.6}
\end{eqnarray} The Floquet index in~(\ref{3.4a}) is thus
\begin{eqnarray}
\alpha_r &=& -\int_{-T}^T d\tau E_r(\tau) - \gamma_r(-T\to T). \label{3.6a}
\end{eqnarray}

Given the zero modes above, it can be shown by direct substitution that the
function
\begin{eqnarray} \psi_{n,r}({\bf x},t) &=& \exp \left\{ -i(\omega_n
- \alpha_r)(t+T)/2T \right\} f_r({\bf x},t) \label{3.7} \end{eqnarray} is the
eigenfunction of the Dirac operator
\begin{eqnarray} {\cal D}\psi_{n,r}({\bf
x},t) &=&\lambda_{n,r}\psi_{n,r}({\bf x},t),\nonumber\\ \lambda_{n,r} &=&
\left(\omega_n + \int_{-T}^T d\tau E_r(\tau)
+\gamma_r(-T\to T)\right)/2T. \label{eigen} \end{eqnarray} The boundary
condition to be imposed on $\psi_{n,r}({\bf x},t)$ at $t=-T,T$ is the
anti-periodic condition~\cite{floquet,niemi},
\begin{eqnarray}
\psi_{n,r}({\bf x},T) &=& -\psi_{n,r}({\bf x},-T). \label{bc} \end{eqnarray}
This condition demands that
\begin{eqnarray} \omega_n &=& (2n+1)\pi.
\label{3.8} \end{eqnarray} In anti-commuting $\gamma_5$ with the operator
$\gamma_0\cal D$, which has the same spectrum as $\cal D$, it is seen that
the spectrum contains oppositely signed pairs, $\{\lambda,-\lambda\}$.

We are now in a position to show the anomaly cancellation. Let ${\cal
Z}_D[A]$ be the vacuum-to-vacuum amplitude of one species of
massless Dirac fermions under the presence of external gauge fields $A_i$;
and ${\cal Z}_W[A]$ that of Weyl fermions.

Using the holomorphic representation to derive the path integral as
in~\cite{tdk-berry}, we arrive at
\begin{eqnarray} {\cal Z}_D[A] &=&
\exp\left\{i\sum_{E_r> 0}\gamma_r(-\infty\to\infty) \right\}\left(\prod_{n,r}
\lambda_{n,r}\right),\nonumber\\ &=&
\exp\left\{-i\sum_{E_r<0}\gamma_r(-\infty\to\infty) \right\}\left(\prod_{n,r}
\lambda_{n,r}\right),\nonumber\\ &=& \exp\left\{\frac{i}{2}\sum_r {\rm
sgn}(E_r)\gamma_r(-\infty\to\infty) \right\} \left(\prod_{n,r}
\lambda_{n,r}\right), \label{3.9} \end{eqnarray} up to irrelevant
multiplicative constant.  We have taken the limit $T\to\infty$, and at
infinite times the interaction is switched off so the time dependency is that
of free theory. The product of eigenvalues is the determinant of the Dirac
operator and the exponential contains the Berry's phase of one-particle
equation of motion.  This factor is the feature distinguishes our
quantisation from the text-book approach, which only has the product of
eigenvalues.

The factor one-half in the exponential argument of the last expression is due
to the boundary condition~(\ref{bc}) and is in agreement with~\cite{niemi}.
In~\cite{tdk-berry}, we have not imposed the anti-periodic boundary condition
but functionally integrated over the {\it independent} fermionic fields at
$t=-T,T$.  Here, we wish to employ the mathematical framework
of~\cite{floquet,niemi} and will thus stick to the anti-periodic boundary
condition from now on. The first expression of~(\ref{3.9}) is obtained if we
only integrate over $\psi(T)$; the second is from the integration of only
$\psi(-T)$; and the third is a square root of the product of the first two.

It is obvious, and can be shown rigorously, that
\begin{eqnarray}
{\cal Z}_W[A]
&=& \sqrt{{\cal Z}_D[A]}. \label{3.10} \end{eqnarray} With the
choice of the square root of the product of $\lambda$'s as the product of all
$\lambda$'s corresponding to positive $E_r$, say,
\begin{eqnarray}
\sqrt{\prod_{n,r} \lambda_{n,r}} &\to& \prod_{E_r>0} \lambda_{n,r},
\label{3.10b} \end{eqnarray} we can then employ the first expression
of~(\ref{3.9}) to have
\begin{eqnarray} {\cal Z}_W[A] &=& \prod_{E_r>0}
{\left(\exp{\left\{\frac{i}{2}
\gamma_r(-\infty\to\infty)\right\}}\lambda_{n,r}\right)}. \label{3.10c}
\end{eqnarray} This expression is sufficient for our present purpose but in
general one can always write ${\cal Z}_W$ as a product over modes singly
taken from oppositely signed pairs
\begin{eqnarray} {\cal Z}_W[A] &=&
\prod_{n,r}{}' {\left(\exp{\left\{\frac{i}{2}{\rm sgn}(E_r)
\gamma_r(-\infty\to\infty)\right\}}\lambda_{n,r}\right)}. \label{3.10a}
\end{eqnarray} That is, the product $\prod{}'$ is over half of as many
eigenmodes of the Dirac operator, none of which belong to the same pair of
oppositely signed eigenvalues.

We digress here for a derivation of~(\ref{3.10a}).  Firstly, by the
substitution $t\to -t$ in the Dirac equation for the zero mode corresponding
to $-E_r$, we have
\begin{eqnarray} f_{(-E_r)}(-t) &=& f_{E_r}(t). \label{1}
\end{eqnarray} From the defintion of the floquet index for $f_{(-E_r)}(t)$,
\begin{eqnarray} f_{(-E_r)}(T) &=& \exp\{-i\alpha_{(-E_r)}\}f_{(-E_r)}(-T),
\label{2} \end{eqnarray} it thus follows that
\begin{eqnarray} f_{E_r}(-T)
&=& \exp\{-i\alpha_{(-E_r)}\}f_{E_r}(T). \label{3} \end{eqnarray} Adirect
comparison with~(\ref{3.4a}) yields
\begin{eqnarray} \alpha_{E_r} &=&
-\alpha_{(-E_r)},{\rm  \;\;\;(mod } \;\;2\pi), \label{4} \end{eqnarray}
from which
\begin{eqnarray} \gamma_{E_r}(-\infty\to\infty) &=&
-\gamma_{(-E_r)}(-\infty\to\infty), {\rm  \;\;\;(mod } \;\;2\pi).
\label{3.9a}
\end{eqnarray} From this we see that, with $\chi_{n,r} \equiv
\exp{\left\{\frac{i}{2}{\rm sgn}(E_r)
\gamma_r(-\infty\to\infty)\right\}}\lambda_{n,r}$,
\begin{eqnarray} {\cal
Z}_D[A] &=& \prod_{n,r}\chi_{n,r} \label{5} \end{eqnarray} consists of, up to
irrelevant multiplicative constant, the pairs $\{\chi,-\chi\}$, resulting
in~(\ref{3.10a}). (The result~(\ref{3.9a}) also confirms the equality of the
first and second expressions of~(\ref{3.9}).)

Let us come back to the main theme and introduce a fifth coordinate, $t_5\in
[0,1]$, to extrapolate between the original and the transformed potentials
under a topologically non-trivial gauge transformation, $U({\bf x})$, which
is time-independent in the temporal gauge,
\begin{eqnarray} A_i(t_5) &=&
(1-t_5)A_i({\bf x},t) + t_5 A_i^U({\bf x},t);\\ A_i^U({\bf x},t) &=& U({\bf
x})A_i({\bf x},t)U^\dagger({\bf x}) + iU({\bf x})\partial_i U^\dagger({\bf
x}). \nonumber \label{3.11} \end{eqnarray} At $t_5 = 0$ and $t_5 = 1$, the
spectra of the Dirac operator are identical.  But in the passage in between,
there is an odd number of pairs of eigenvalues, denoted by $\lambda^*$,
crossing zero as follows from a Wick rotation of Witten's results for the
euclidean version of the operator~(\ref{3.0}),
\begin{eqnarray}
\lambda^*_{n,r}(t_5=1) &=& -\lambda^*_{n,r}(t_5=0). \label{3.12}
\end{eqnarray} The sign of~(\ref{3.10c},\ref{3.10a}) would have switched and
resulted in the Witten anomalies for the Weyl fermions, were the accompanying
exponential factor not there, as we shall see.

{}From the gauge covariance of the Dirac operator,
\begin{eqnarray} {\cal
D}(t_5=1) &=& U{\cal D}(t_5=0) U^\dagger, \label{3.13} \end{eqnarray} it
follows that
\begin{eqnarray} \psi_{\lambda^*(t_5=1)}(t_5 = 1) &=&
U\psi_{-\lambda^*(t_5=0)}(t_5 = 0). \label{3.14} \end{eqnarray} For these
crossing modes, we also have
\begin{eqnarray} E^*_r(t_5=1) &=& -E^*_r(t_5=0)
\label{3.15} \end{eqnarray} because of the gauge covariance of $H(t)$,
\begin{eqnarray} E^*_r(t_5=1)\psi_{\lambda^*(t_5=1)}(t_5=1) &=&
H(t_5=1)\psi_{\lambda^*(t_5=1)},\nonumber\\ &=&
\left(UH(t_5=0)U^\dagger\right) \left(U\psi_{-\lambda^*(t_5=0)}(t_5 =
0)\right),\nonumber\\ &=& -E^*_r(t_5=0)U\psi_{-\lambda^*(t_5=0)}(t_5 =
0),\nonumber\\ &=& -E^*_r(t_5=0)\psi_{\lambda^*(t_5=1)}(t_5=1).\nonumber
\label{3.16} \end{eqnarray} Substituting~(\ref{eigen},\ref{3.12})
into~(\ref{3.12}), we
can derive the gauge transformed of the corresponding Berry's phase
\begin{eqnarray}
\gamma^*_r(t_5=1)
&=& -2\omega_n - \gamma^*_r(t_5=0),\label{3.17} \end{eqnarray}
where we have employed the fact that $\omega_n$ is independent of the gauge
transformations because of the independence of the anti-periodic boundary
condition~(\ref{bc}). The gauge transformation of the exponential argument
in~(\ref{3.10c},\ref{3.10a}) is then
\begin{eqnarray} \left[\frac{i}{2}{\rm
sgn}(E_{r}^*)\gamma_{r}^*\right]_{t_5=1} &=& \left[ \frac{i}{2}{\rm
sgn}(E_{r}^*)\gamma_{r}^*\right]_{t_5=0} + i{\rm sgn}\left(E^*_r(t_5=0)
\right)\omega_n. \label{3.18} \end{eqnarray}

For the other eigenmodes with $\lambda$ not crossing zero as $t_5$ is varied
between $0$ and $1$, there is no sign change for $\lambda$ and $E_r$ at
$t_5=0,1$.  And the corresponding Berry's phase also remains the same.

Thus for each factor $(-1)$ from an eigenvalue of the Dirac operator changing
sign, we pick up {\em another} corresponding factor
\begin{eqnarray} \exp(\pm
i \omega_n) &=& (-1) \label{3.19} \end{eqnarray} from the
accompanying exponential in~(\ref{3.10c},\ref{3.10a}). The generating path
integral of chiral fermions is consequentially gauge invariant and has no
anomalies associated with large gauge transformations,
\begin{eqnarray} {\cal
Z}_W[A^U] &=& {\cal Z}_W[A].\label{3.20} \end{eqnarray}

In summary, with the Berry's phase needed to satisfy certain boundary
conditions of the generating path integral, the gauge anomalies are shown to
be eliminated even for an odd number of chiral fermions, without any extra
quantum fields introduced.  This is in contra-distinction to other
quantisations of `anomalous' gauge theory where extra, new fields are
introduced to explicitly cancel the anomalies~\cite{others}.  In the
cancellation itself, there is a further point which distinguishes these
approaches from ours.  In ours, the cancellation of the anomalies is only the
{\it consequence} of a {\it consistent} approach to quantisation, either with
the Berry's phase or with a careful reconsideration of the time-ordered
product~\cite{tdk-ordered}.  On the other hand, it is still unclear whether
the {\it ad hoc} addtion of fields with the sole purpose to cancel the
anomalies is equivlent to our approach or not; and, more importantly, whether
such addition would uphold the consistency, of the path integral or the
Tomonaga-Schwinger equation, which is the starting point of our approach.

The cancellation of the Witten anomalies and the absence of the
Adler-Bell-Jackiw gauge anomalies in the hamiltonian approach is addressed
elsewhere~\cite{tdk-inv}. Our result of the cancellation is in no way
diminishing the importance of the Witten or the ABJ gauge anomalies.  With
hindsight, they are invaluable in exposing the shortcomings of the
conventional quantisation of field theory.

I am indebted to the moral support over the years from Brian Pendleton, David
Wallace, Dick Dalitz, Ian Aitchison, Michael Danos and Bruce McKellar.  I
also wish to acknowledge the financial support of the  Australian Research
Council.
 \end{document}